\documentclass[prl,aps,reprint,notitlepage,showpacs,twocolumn,letterpaper,nofootinbib,longbibliography]{revtex4-1}
\pdfoutput=1
\usepackage{epsfig}
\usepackage{amsfonts}
\usepackage{amsmath}
\usepackage{graphicx}
\usepackage{color}
\usepackage{amssymb,amsmath,hyperref,slashed,color,graphicx,bm,cancel,url}
\usepackage[T1]{fontenc}
\usepackage{relsize}
\hypersetup{colorlinks,citecolor= red,linkcolor= niceblue, urlcolor=niceblue}
\definecolor{niceblue}{rgb}{0.1,0.2,0.6}
\def\Carleton{Department of Physics, Carleton University, Ottawa, K1S 5B6, Canada}

\begin{document}

\title{Higgs Portal From The Atmosphere To Hyper-K}

\author{Paul Archer-Smith}
\affiliation{\Carleton}
\author{Yue Zhang\vspace{0.1in}}
\affiliation{\Carleton}
\date{\today}

\begin{abstract}
A light Higgs portal scalar could be abundantly produced in the earth's atmosphere and decay in large-volume neutrino detectors. 
We point out that the Hyper-Kamiokande detector bears a strong discovery potential of probing such particles in an uncharted parameter space that is  actively explored by intensity frontier experiments including rare kaon decays.
The signal we propose to look for is electron-positron pair creation that manifests as a double-ring appearing from the same vertex. Most of these pairs originate from zenith angles above the Hyper-K detector's horizon. This search can be generalized to other new light states and is highly complementary to beam experiments.
\end{abstract}

\maketitle 

A Standard Model gauge singlet scalar that mixes with the Higgs boson, sometimes also referred to as the ``dark Higgs'', is a simple new physics candidate.
It has been introduced for exploring the dark universe~\cite{Patt:2006fw, Weinberg:2013kea, Wise:2014jva, Wise:2014ola, Zhang:2015era}, facilitating baryogengesis mechanisms~\cite{Anderson:1991zb, Pietroni:1992in, Carena:2018cjh, Carena:2019xrr}, precision physics of the Standard Model~\cite{TuckerSmith:2010ra, Chen:2015vqy}, and, perhaps, naturalness~\cite{Graham:2015cka}. 
In its minimal incarnation, the Higgs portal scalar is produced in laboratories and decays into Standard Model particles via the same mixing parameter with the Higgs boson. These makes it a well-motivated and well-defined target of searches in a number of experiments. Constraints have been set for a wide range of its mass~\cite{Beacham:2019nyx, Flacke:2016szy, Clarke:2013aya}. 
In particular, if the scalar is lighter than $\sim$ GeV, leading constraints come from the measurement of rare $K$ and $B$ meson decays where the mixing parameter must be smaller than $\sim10^{-3}$.

Recently, Higgs portal scalar has been revisited for understanding a new experimental finding. In 2016-18, the KOTO experiment at J-PARC performed a search for the flavor-changing decay process $K_L \to \pi^0 \nu\bar\nu$, in final states with two energetic photons plus a missing transverse momentum. It was originally reported that four candidate events were identified whereas the Standard Model predicts nearly none~\cite{KOTOConfer1, KOTOConfer2, KOTO}. 
Although in a more recent analysis by the KOTO collaboration, the significance of this excess is substantially reduced~\cite{Ahn:2020opg},
it has triggered the exploration of a variety of potential new physics behind, heavy and light.
Among them, a light Higgs portal scalar $\phi$ stands out as the simplest candidate~\cite{Egana-Ugrinovic:2019wzj} (see also~\cite{Kitahara:2019lws, Dev:2019hho, Liu:2020qgx}). 
The signal can be explained as $K_L \to \pi^0 \phi$ decay where $\phi$ is long lived and escapes the detector. 
The potentially relevant parameter space, corresponding to a $\phi$ mass between 100-200 MeV,
and $\phi$-Higgs mixing parameter of (a few) $\times\, 10^{-4}$,
is a blindspot of existing searches at the intensity frontier.
The most direct cross check is the isospin related decay mode, $K^+\to \pi^+ \phi \to \pi^+ + {\rm invisible}$, which can be constrained by the $K^+\to \pi^+ \nu\bar\nu$ measurement. Indeed, this channel has been searched for at the E949~\cite{Artamonov:2009sz} and NA62~\cite{CortinaGil:2020vlo, CortinaGil:2020zwa, CortinaGil:2020fcx} experiments where upper limits are set on the mixing parameter of the Higgs portal scalar. However, both limits feature a gap when the scalar mass is around the pion mass, due to the enormous $K^+\to \pi^+\pi^0$ background. 
In this mass window, the best upper limit on the mixing parameter is set by an early beam dump experiment, CHARM~\cite{Bergsma:1985qz}, in the search for displaced decay of $\phi$.

This comparison points to a direction to proceed. In order to cover the Higgs portal scalar in the above blindspot, one should resort to appearance experiments hunting the visible decay of long lived $\phi$ particles rather than disappearance experiments searching for $\phi$ as missing momentum.
As a further useful observation, with a mixing with Higgs boson $\sim 10^{-4}$, the decay length of Higgs portal scalar is of order hundreds of kilometers, and even longer if boosted. This gives motivation to imagine large experiments operating at length scales beyond those beam-based ones built entirely within the laboratories.

In this Letter, we propose using a nature-made experimental setup to probe the Higgs portal scalar $\phi$. It utilizes cosmic rays as the beam, earth's atmosphere as the target, and earth itself as the shielding region.
In this picture, $\phi$ particles originate from the decay of kaons, with the latter being abundantly produced in the cosmic-ray-atmosphere fixed-target collisions, together with charged pions that make the atmospheric neutrinos~\cite{Fukuda:1998mi}. If long lived enough, the $\phi$ particles travel a long distance across the earth before decaying inside a human-made detector.
We focus on the Hyper-Kamiokande (Hyper-K) experiment \cite{Hyper-Kv2} which, at least for the foreseeable future, has the largest detector volume and a suitably low energy threshold to capture the scalar decays.

The Higgs portal scalar is defined as a mass eigenstate and a linear combination of a Standard Model gauge singlet $s$ and the Higgs boson $h$,
\begin{equation}
\phi = \cos\theta\, s + \sin\theta\, h \ ,
\end{equation}
where $\theta$ is a real mixing parameter. Such a singlet-Higgs mixing can be generated by adding to the standard model Lagrangian a Higgs portal interaction term $\mu s H^\dagger H$. After the electroweak symmetry breaking, it generates a bilinear terms that allows the singlet scalar to mix with the Higgs boson. This corresponds to the minimal scenario of scalar portal (BC4) considered in the community report Ref.~\cite{Beacham:2019nyx}.
For simplicity, we proceed the following discussions using the above phenomenological parametrization.

\begin{figure}[t]
\centerline{\includegraphics[width=0.45\textwidth]{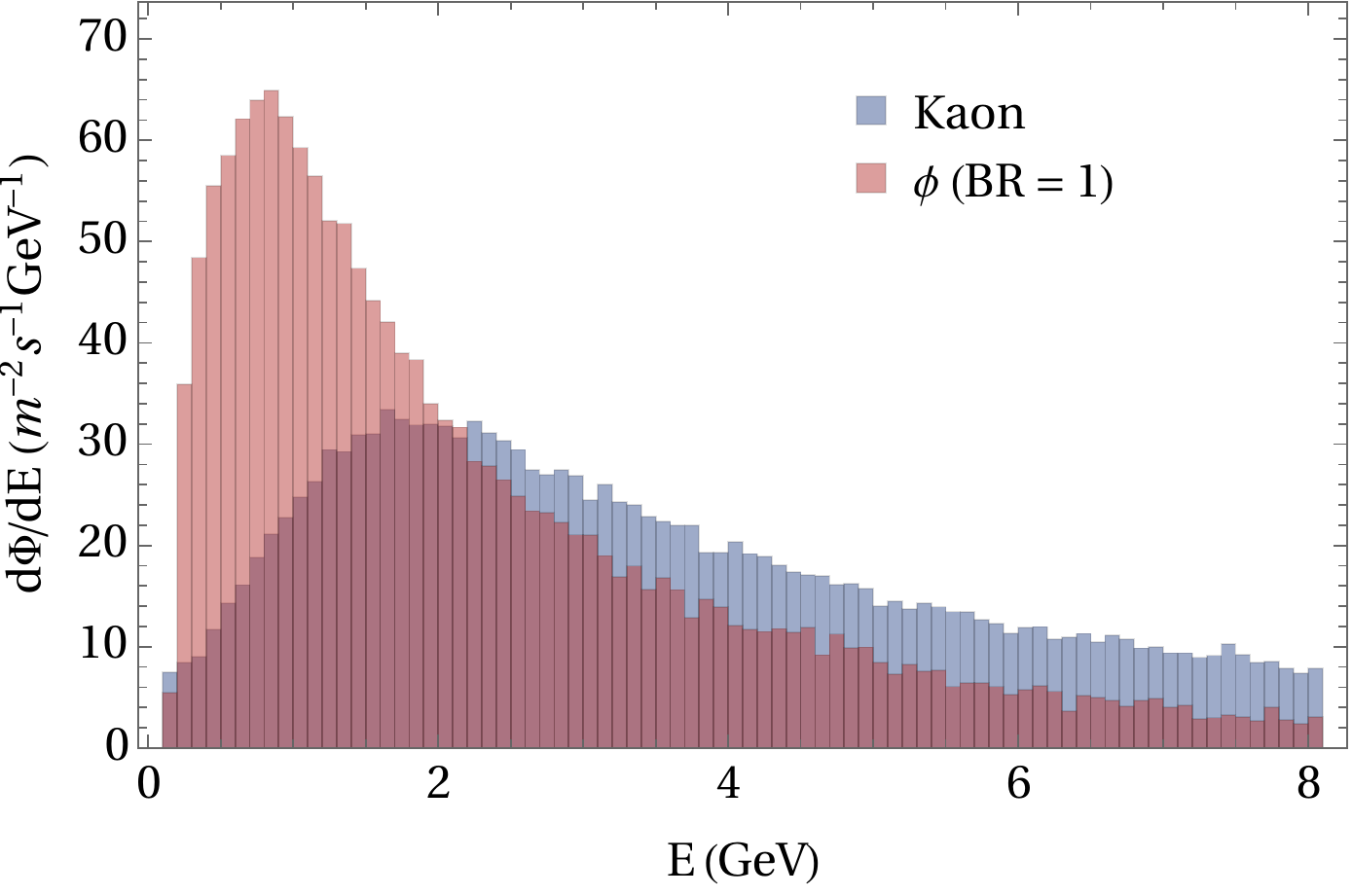}}
\caption{Energy distribution of atmospheric kaons ($K^\pm$ and $K_L$ added together) and $\phi$ particles, for $m_\phi=150\,$MeV, obtained from the atmospheric simulation described in the text. 
For illustration purpose, the flux of $\phi$ has been rescaled by assuming the $K\to\pi\phi$ decay branching ratios are equal to 1.
}\label{fig:fluxes}
\end{figure}

The cosmic rays near us are dominated by protons while the elements in the earth's atmosphere are dominated by nitrogen and oxygen, comprised of equal numbers of protons and neutrons.
We simulate fixed target proton-proton and proton-neutron collisions using {\tt PYTHIA 8}~\cite{Sjostrand:2014zea} for various incoming proton energies,
which is further convoluted with the incoming cosmic proton spectrum~\cite{Tanabashi:2018oca} to derive 
the differential energy spectrum of kaons (most relevant for this study, $K^\pm$ and $K_L$), ${d \Phi}/{dE_K}$. 
Their sum is shown as the blue histogram in Fig.~\ref{fig:fluxes}.
The ratio of $K^\pm$ and $K_L$ particles is about $2:1$, as expected.

The $\phi$ particles are produced from rare kaon decays, $K^\pm\to \pi^\pm \phi$ and $K_L \to \pi^0 \phi$. The corresponding branching ratios are~\cite{LEUTWYLER1990369,Feng:2017vli, Batell:2019nwo, Gunion:1989we}
\begin{eqnarray}\label{Kpm}
&&{\rm Br}(K^\pm\to \pi^\pm \phi) \simeq \frac{9 \tau_{K^\pm} |V_{ts} V_{td}^*|^2 G_F^3 m_t^4 m_{K^\pm}^2 p_{\phi\rm CM}  \theta^2}{2048\sqrt{2} \pi^5},\\
&&{\rm Br}(K_L \to \pi^0 \phi) \simeq \frac{9 \tau_{K_L} [{\rm Re}(V_{ts} V_{td}^*)]^2 G_F^3 m_t^4 m_{K^\pm}^2 p_{\phi\rm CM}  \theta^2}{2048\sqrt{2} \pi^5},\nonumber
\end{eqnarray}
where the decay momentum in the center-of-mass (CM) frame is $p_{\phi\rm CM} = \lambda(m_K^2, m_\pi^2, m_\phi^2)/2m_{K^\pm}$, and $\lambda$ is the K\"all\'en function.
In small $m_\phi$ limit, ${\rm Br}(K_L \to \pi^0 \phi)/{\rm Br}(K^\pm\to \pi^\pm \phi) \simeq 3.7$~\cite{Grossman:1997sk}.
In the lab frame, the ratio of the final state $\phi$ energy to that of kaon is
\begin{equation}\label{EphiEK}
\frac{E_\phi}{E_K} =
\frac{E_{\phi\rm CM}}{m_K} + \frac{p_{\phi\rm CM}}{m_K} \sqrt{ 1 - \frac{m_K^2}{E_K^2}} \cos\vartheta_{\rm CM} \ ,
\end{equation}
where $E_{\phi\rm CM}=\sqrt{p_{\phi\rm CM}^2 + m_\phi^2}$ and $\vartheta_{\rm CM}$ is the relative angle between $\phi$'s three-momentum in the kaon rest frame and the boost direction of the kaon.
Because $K^\pm$ and $K_L$ are scalars, the angular $\phi$ distribution in their rest frame is isotropic.
For given energy $E_K$, the values of $E_\phi$ distribute evenly between its extremes, corresponding to $\cos\vartheta_{\rm CM}=\pm1$.
The resulting differential flux of $\phi$ can be calculated using
\begin{eqnarray}\label{dPhiphi}
\frac{d \Phi_\phi}{dE_\phi} &=& \sum_{K= K^\pm, K_L}  {\rm Br}(K\to \pi \phi) \int_{E_{K\rm min} (E_\phi)}^{E_{K\rm max} (E_\phi)} dE_K  \frac{d \Phi_K}{dE_K} \nonumber\\
&&\hspace{1cm}\times \frac{m_K}{2 p_{\phi\rm CM}\sqrt{E_K^2-m_K^2}} \ ,
\end{eqnarray}
where $E_{K\rm max, min}$ is the largest (smallest) kaon energy that satisfies Eq.~(\ref{EphiEK}), for given $E_\phi$.
In the limit $E_K\gg m_K$, $E_{K\rm max, min}\simeq E_\phi m_K/(E_{\phi\rm CM} \mp p_{\phi\rm CM})$.
In Fig.~\ref{fig:fluxes}, the red histogram shows the energy distribution of atmospheric $\phi$ particles, for $m_\phi=150\,$MeV.
Its energy is peaked $\sim700$\,MeV.

It is worth pointing out that when simulating atmospheric $\phi$ production, we have we restrict the CM energy of $pp$ and $pn$ scatterings to be above $\sim 6$\,GeV in order for the parton picture used by {\tt PYHTIA} to be valid.
We also neglected secondary reactions of the produced hadrons with the atmosphere before they decay, keeping in mind that the earth's atmosphere is dilute. 
Both processes could in principle result in more kaon (and thus $\phi$ particles) production. 
We made the above approximations for simplicity.

After being produced in the atmosphere, the $\phi$ particles can travel through the earth to decay inside human-made detectors, provided they have sufficiently long lifetimes. Clearly, the larger the detector the better to capture such a signal. Its energy threshold should be low enough to see sub-GeV energy deposits from the $\phi$ decay. These requirements led us to consider Hyper-K.

\begin{figure}[t]
\centerline{\includegraphics[width=0.4\textwidth]{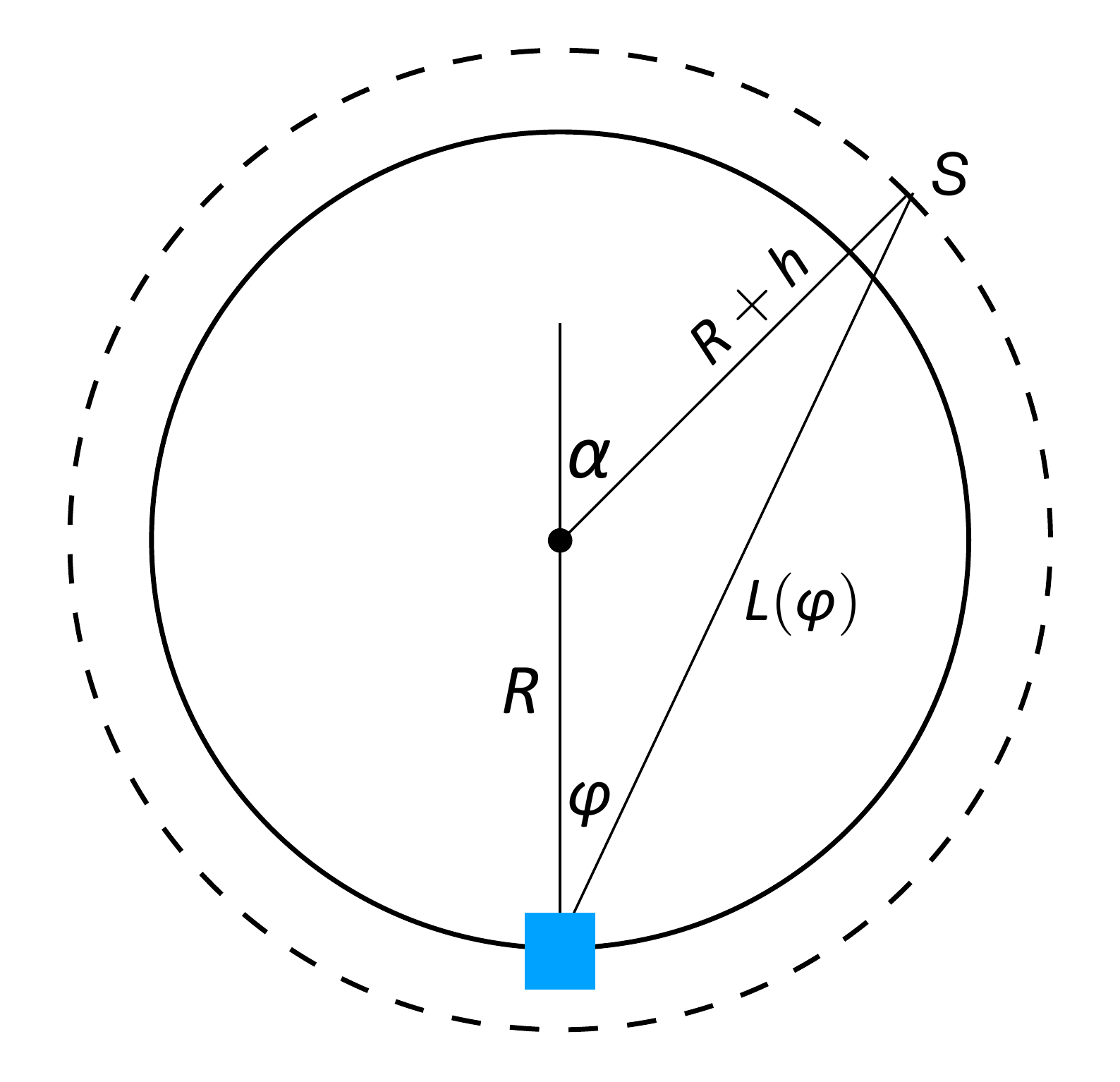}}
\caption{Geography of earth and detector. The blue box indicates the location of the Hyper-K detector. The 
dashed circle represents a sphere where the cosmic-ray-atmosphere 
reactions mainly occur that produce light $\phi$ particles. 
$h$ is given by the height of this sphere plus the 
depth of detector underground, and $\varphi$ is the zenith 
angle in view of the detector.}\label{geometry}
\end{figure}

\begin{figure}[p]
\centerline{\includegraphics[width=0.45\textwidth]{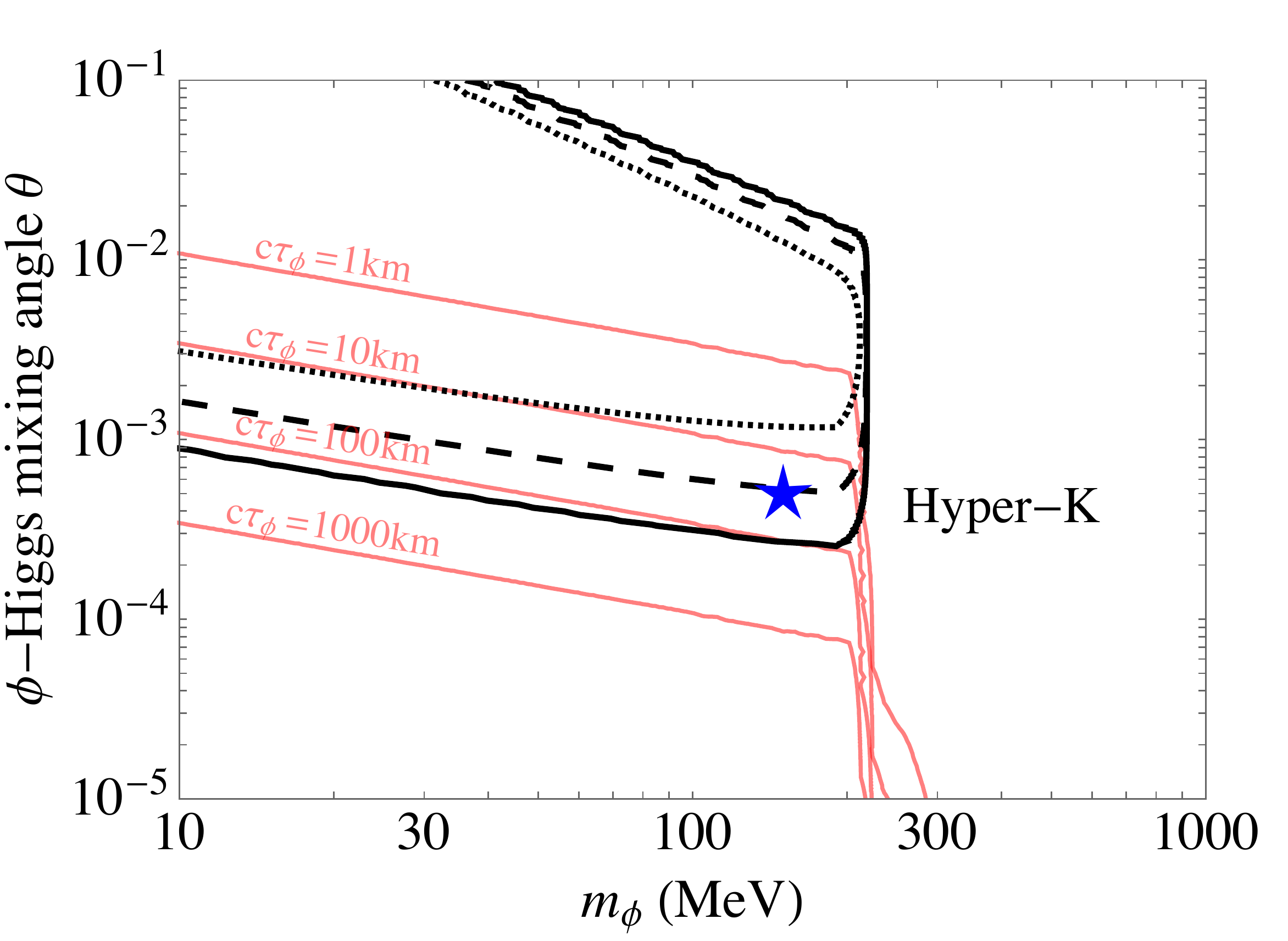}}
\centerline{\includegraphics[width=0.45\textwidth]{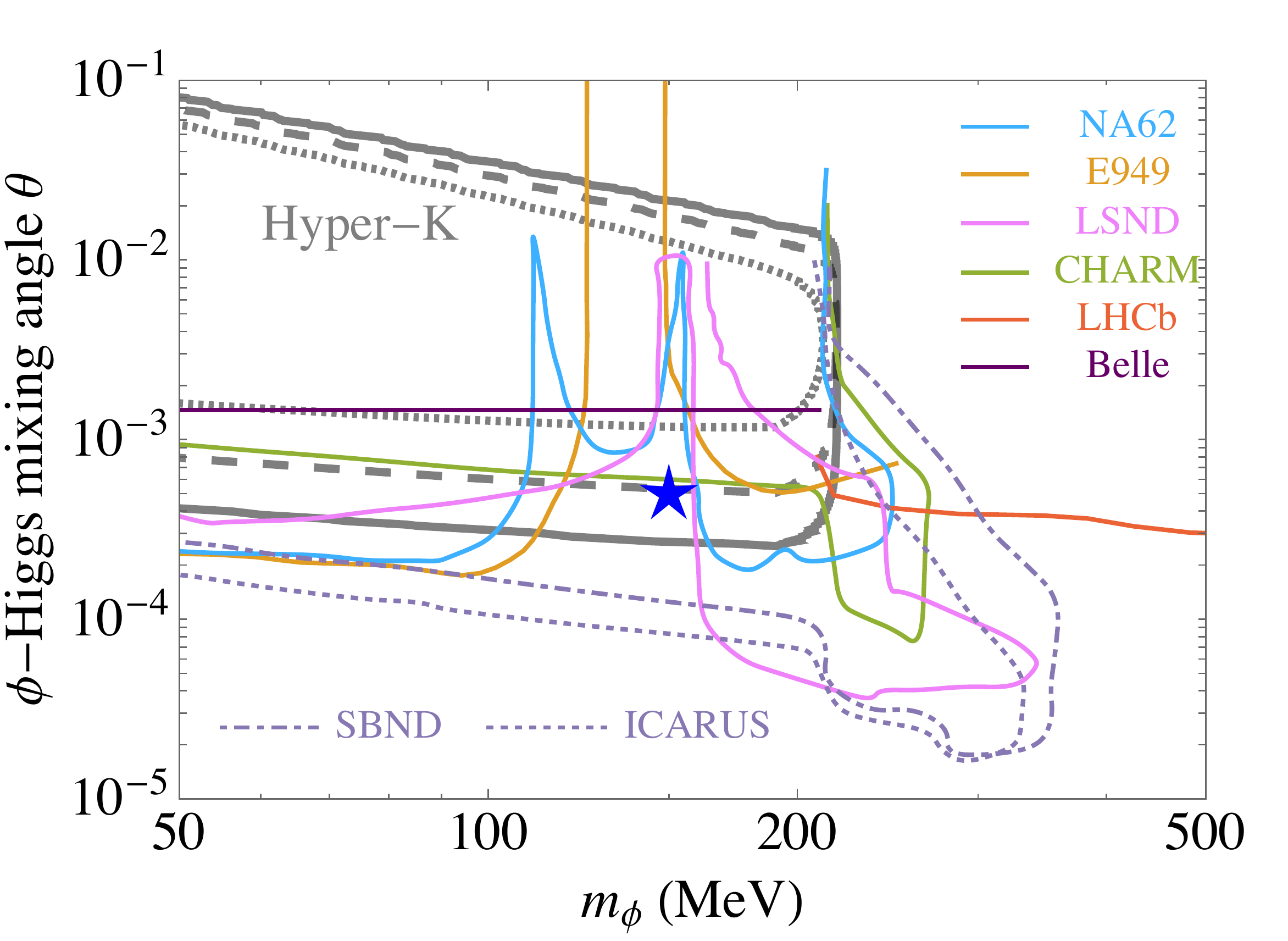}}
\centerline{\includegraphics[width=0.45\textwidth]{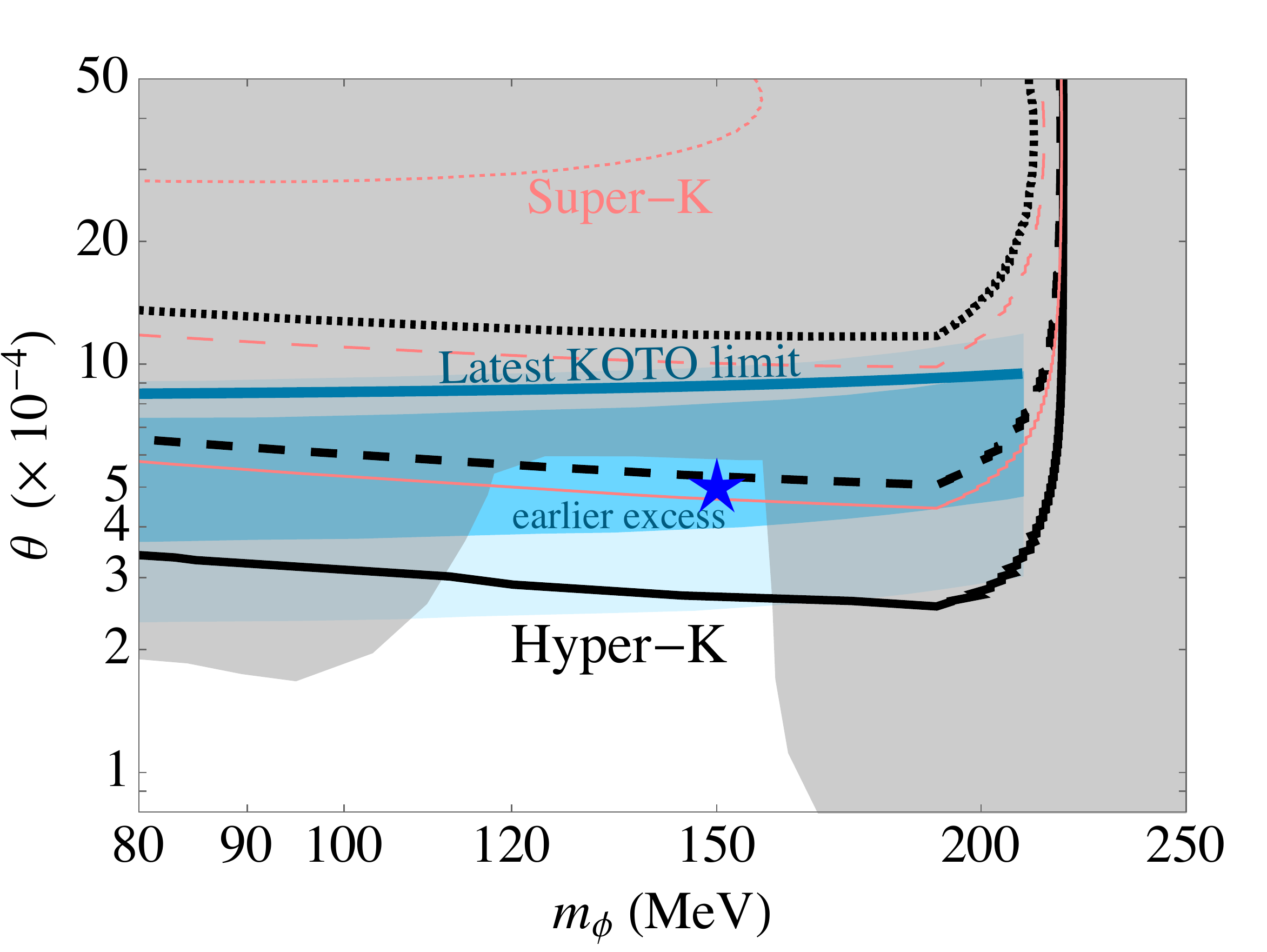}}
\caption{\scriptsize
{\bf Upper:} Using Hyper-K detector to search for long lived Higgs portal scalar $\phi$ produced from the atmosphere. The (solid, dashed, dotted) black contours correspond to 10, 100, 1000 signal events after ten years of exposure. The red curves corresponds to constant values of $c\tau_\phi$, the lifetime of  $\phi$ times the speed of light. 
{\bf Middle}: Hyper-K region (the three gray contours in the background are same as those black ones in the upper panel) shown in together with the existing constraints (from E949, NA62, CHARM, LSND, LHCb, Belle) and future reach by the upcoming experiments (ICARUS, SBND).
{\bf Lower:} The region of parameter space favored by the earlier KOTO excess is shown by the blue bands (dark and light blue correspond to 1 and $2\sigma$ favored regions, respectively)~\cite{KOTOConfer1, KOTOConfer2, KOTO}. 
The upper bound on the mixing angle $\theta$ derived from the latest KOTO analysis~\cite{Ahn:2020opg} is down by the thick dark blue curve.
The union of existing constraints exclude the gray shaded region. Like the upper and middle panels, the black curves corresponds to fixed number of signal events using Hyper-K to hunt atmospheric $\phi$ particles. For comparison, we also show the rescaled 10, 100, 1000 signal event contours (in pink color) for Super-K with 328 kiloton-year of data.
In all the plots, the blue fivestar corresponds to the benchmark point used in Eq.~\eqref{decaylength} and Fig.~\ref{fig:kinematics}.
}\label{mainplot}
\end{figure}

To calculate the $\phi$ flux at Hyper-K detector, we consider the geometric picture shown in Fig.~\ref{geometry}.
We assume all cosmic-ray-atmosphere reactions occur on a sphere with fixed height above the ground. This height plus 
the depth of the underground Hyper-K detector, denoted by $h$, is taken to be 10\,km.
The angles $\varphi$ and $\alpha$ are related by
\begin{eqnarray}
\cos\alpha= [L(\varphi) \cos\varphi - R]/(R+h) \ , 
\end{eqnarray}
where $L(\varphi)$ is the distance $\phi$ travels,
\begin{eqnarray}\label{Lvarphi}
L(\varphi) = R \cos\varphi + \sqrt{h^2 + 2 R h + R^2 \cos^2\varphi} \ .
\end{eqnarray}
An infinitesimal area on the source sphere is
\begin{equation}
d \mathcal{S} = 2\pi (R+h)^2 d\cos\alpha= \frac{2 \pi (R+h)L(\varphi)^2}{L(\varphi) - R\cos\varphi}  d \cos\varphi \ . 
\end{equation}
We assume cosmic ray showers on the earth atmosphere to be isotropic, and so is the resulting $\phi$ angular distribution within the hemisphere pointing towards the center of the earth.~\footnote{It is a simplification that we assumed the atmospheric secondaries are produced isotropically inwards to the earth, which allows us to proceed the flux calculation analytically. We expect this approximation to be appropriate because we consider the high energy cosmic ray collisions (with center of mass energy above 6 GeV). As a result, the kaons and subsequently the $\phi$ particles are produced along the forward direction of the scatterings. In order for the $\phi$ particle to reach Hyper-K which is underground, the original cosmic ray direction cannot be tangential to the upper atmosphere.}
The flux of long-lived $\phi$ at the detector is related to its flux at the source (atmosphere) by a geometric factor,
\begin{equation}
\left(\frac{d \Phi_a}{dE_a}\right)_{\rm detector} =  \left(\frac{d \Phi_a}{dE_a}\right)_{\rm source} \int \frac{d \mathcal{S}}{2\pi L(\varphi)^2} \ ,
\end{equation}
where the latter takes the form
\begin{equation}
\int \frac{d \mathcal{S}}{2\pi L(\varphi)^2} =\int_{0}^\pi  \sin\varphi d \varphi \frac{(R+h)}{L(\varphi) - R\cos\varphi} \ .
\end{equation}
If the Hyper-K detector volume is denoted by $V$, the event rate of long-lived $\phi$ particles decaying inside this volume is, regardless of its shape,
\begin{eqnarray}\label{eq:rate}
R_{\rm event} &=& V \int_{0}^\pi  \sin\varphi d \varphi \frac{R+h}{L(\varphi) - R\cos\varphi} \nonumber \\
&& \hspace{1.1cm }\times \int d E_\phi \frac{d \Phi_\phi/dE_\phi}{\gamma \beta \tau_\phi} e^{- \frac{L(\varphi)}{\gamma \beta \tau_\phi}} \ ,
\end{eqnarray}
where $\gamma$ is the boost factor of $\phi$ with energy $E_\phi$ and $\beta$ is the corresponding velocity. 
$d \Phi_\phi/dE_\phi$ is given by Eq.~(\ref{dPhiphi}).
The lifetime of $\phi$ is dictated by the Higgs portal. For mass of  $\phi$ below twice the muon mass, it mainly decays into a $e^+e^-$ pair. The corresponding decay length without boost factor is (assuming $m_\phi \gg m_e$)
\begin{equation}\label{decaylength}
\begin{split}
c \tau_\phi &= \frac{8\pi}{\sqrt{2} G_F m_e^2 m_\phi \theta^2} \\
& \simeq 30\,{\rm km} \left( \frac{0.15\,{\rm GeV}}{m_\phi} \right) \left( \frac{5\times10^{-4}}{\theta} \right)^2 \ ,
\end{split}
\end{equation}
where the benchmark values of $\theta$ and $m_\phi$ corresponds to the point indicated by the blue fivestar in Fig.~\ref{mainplot}.

It is worth noting that the small electron mass appearing in the decay rate does not suppress the $\phi$ production rate (see Eq.~(\ref{Kpm})).\footnote{This is in sharp contrast with the case of dark photon where the same parameter controls both the production and decay. In fact, Ref.~\cite{Arguelles:2019ziu} found that the atmospheric production cannot provide a competitive constraint for dark photon.}
Once produced from the atmosphere, it is able to penetrate the earth above deep underground detectors. In water Cherenkov detectors like Hyper-K, the final state $e^+e^-$ manifest as a double-ring signature, where the two rings originate from the same primary vertex of $\phi$ decay. 
We focus on fully contained events where the $\phi$ decay vertex emerges from inside the detector.

Our main result is shown in Fig.~\ref{mainplot}, in the $\theta$ versus $m_\phi$ plane.
In the upper panel, the black solid, dashed, and dotted curves corresponds to observing 10, 100, and 1000 $e^+e^-$ pair events due to $\phi$ decay in the Hyper-K detector, after 10 years of data taking.   
To derive these curves, the volume of the Hyper-K detector used is $216\times10^3\,{\rm m^3}$ (diameter = 70.8\,m and height = 54.8\,m)~\cite{Hyper-K}.
Beyond the top and right boundaries of the covered regions (enclosed by the black curves), the $\phi$ particles decays too fast to reach the detector and produce enough signal events, either due to large mixing angle $\theta$ or the opening of the $\phi\to\mu^+\mu^-$ decay channel. The lower boundaries of the covered regions are simply set by the production rate which is proportional to $\theta^2$.
In the same plot,  the red contours correspond to constant values of $c\tau_\phi$, the lifetime of  $\phi$ times the speed of light.
They are not parallel to the upper edges of the black contours because $\phi$ particles are produced boosted.

Here, we only present contours for certain signal events. 
They indicate the region of parameter space that potentially could be covered with the Hyper-K detector.
Once the backgrounds is fully understood, it is straightforward to derive an expected limit using our result.
A thorough background analysis is beyond the scope of this paper.
Potentially important background includes atmospheric neutrinos undergoing neutral-current interaction with a $\pi^0$ radiation. 
The two photons from subsequent $\pi^0$ decay could also manifest as a double ring in water Cherenkov detectors. 
We can make an estimate of it based on a recent Super-Kamiokande analysis~\cite{SuperKResults}.
Number of $\pi^0$-like double ring events has been reported in Fig.~5  and Table II of~\cite{SuperKResults}.
Rescaling the result to Hyper-K, we expect to see a few thousand such background events.
However, it is worth noting that the energy spectrum of the background $\pi^0$ peaks around 200\,MeV whereas the atmospheric $\phi$ energy peaks around 700\,MeV (see Fig.~\ref{fig:fluxes} above). This difference could serve a useful kinematical cut.
In addition, a zenith angle distribution analysis (see Fig.~\ref{fig:kinematics} below) might be useful for further background discrimination.

In the middle panel of Fig.~\ref{mainplot}, we show the above found region of interest to Hyper-K in together with the existing constraints, including the search of $K^\pm\to \pi^\pm \phi \to \pi^\pm + {\rm invisible}$ at E949~\cite{Artamonov:2009sz} (see also \cite{Egana-Ugrinovic:2019wzj}) and the NA62 experiment~\cite{CortinaGil:2020vlo, CortinaGil:2020zwa, CortinaGil:2020fcx}, displaced visibly-decaying $\phi$ search at CHARM~\cite{Bergsma:1985qz, Egana-Ugrinovic:2019wzj} and LSND~\cite{Foroughi-Abari:2020gju}, measurement of $B\to K\phi \to K \mu^+\mu^-$ at LHCb~\cite{Aaij:2016qsm, Aaij:2015tna} and $B\to K \phi \to K + {\rm invisible}$ at Belle~\cite{Chen:2007zk} (see also \cite{Bezrukov:2009yw}). We also show the future reach by the upcoming ICARUS, SBND experiments based on a recent analysis~\cite{Batell:2019nwo}.

In the lower panel of Fig.~\ref{mainplot}, we zoom in toward the parameter space, where $m_\phi \in (100-200)$\,MeV and
$\theta \sim \text{(a few)}\times\, 10^{-4}$, potentially relevant for KOTO.
Again, the Hyper-K coverage is indicated by the thick black curves, with solid, dashed and dotted corresponding to observing 10, 100 and 1000 $e^+e^-$ pair events, respectively.
Remarkably, they cover a new region parameter space that has not been constrained before by any existing experiments. 
For comparison, we also show the rescaled 10, 100, 1000 signal event contours for the Super-Kamiokande experiment with 328 kiloton-year data collection.

Another attenuation effect before $\phi$ reaches the Hyper-K detector is the scattering with the earth. 
The energy of atmospheric $\phi$ is peaked around GeV scale, 
which roughly coincides with the nucleon mass and the QCD scale for strong interactions.
The corresponding scattering cross section of $\phi$ with the nucleon target can be estimated to be, $\sigma_{\phi+N\to\pi^0+N} \sim g_{\pi NN}(\theta^2/m_N)^2 (m_N/v)^2 \sim 10^{-34}\theta^2\,{\rm cm^2}$, where $v$ is the electroweak vaccum expectation value and $m_N$ is the nucleon mass. The extra suppression factor $(m_N/v)^2$ arises from the nucleon-Higgs coupling~\cite{Shifman:1978zn}.
Given the earth nucleon density, $n\sim 10^{24}/{\rm cm^3}$, the free streaming length of $\phi$ is roughly, $l_F =1/(n\sigma) \gtrsim 10^5\, {\rm km} /\theta^2$. 
The free streaming length of $\phi$ through the earth is sufficiently long even for $\theta \sim \mathcal{O}(1)$.

\begin{figure*}
\centerline{\includegraphics[width=0.436\textwidth]{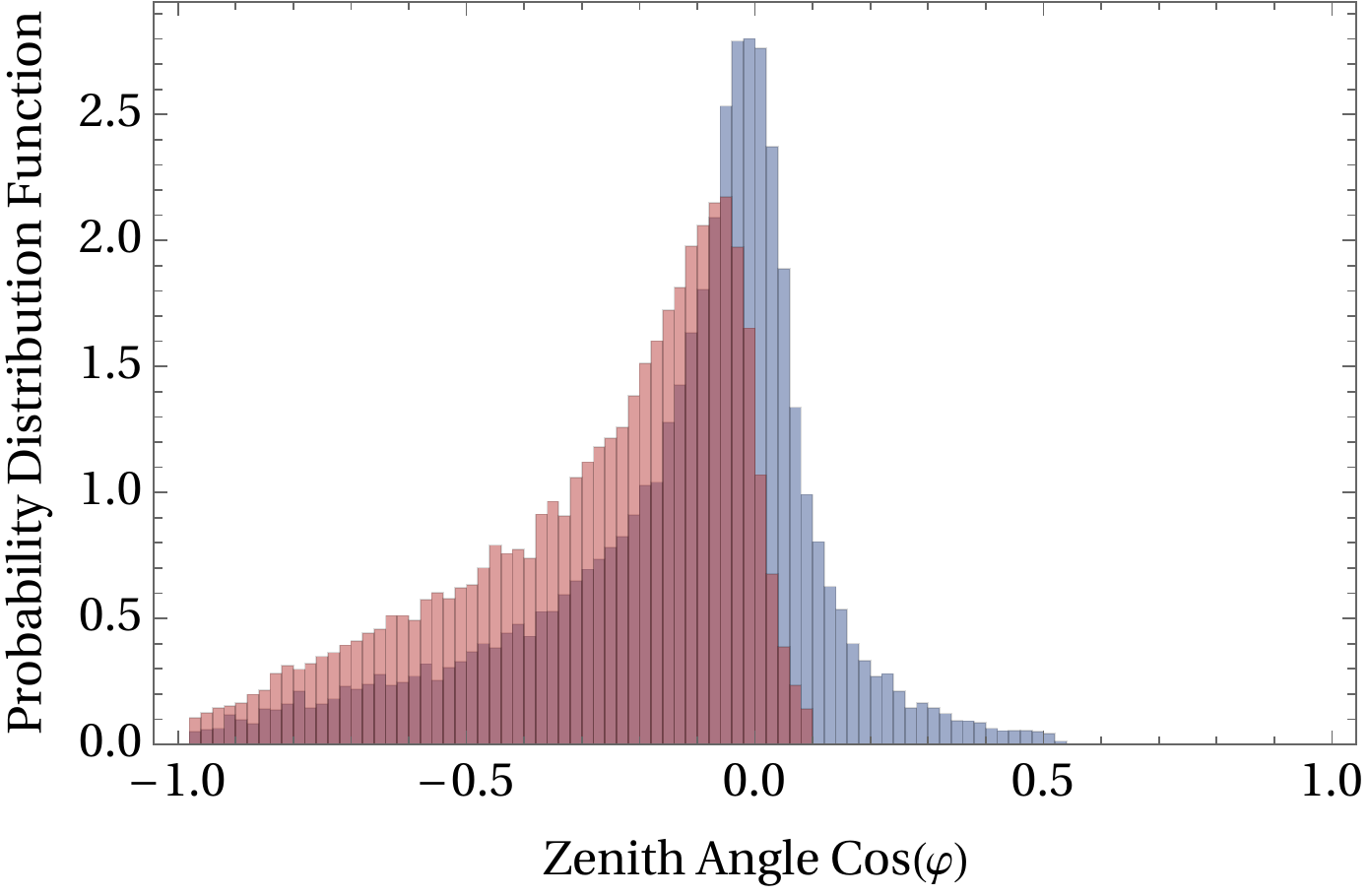}
\hspace{0.5cm}
\includegraphics[width=0.45\textwidth]{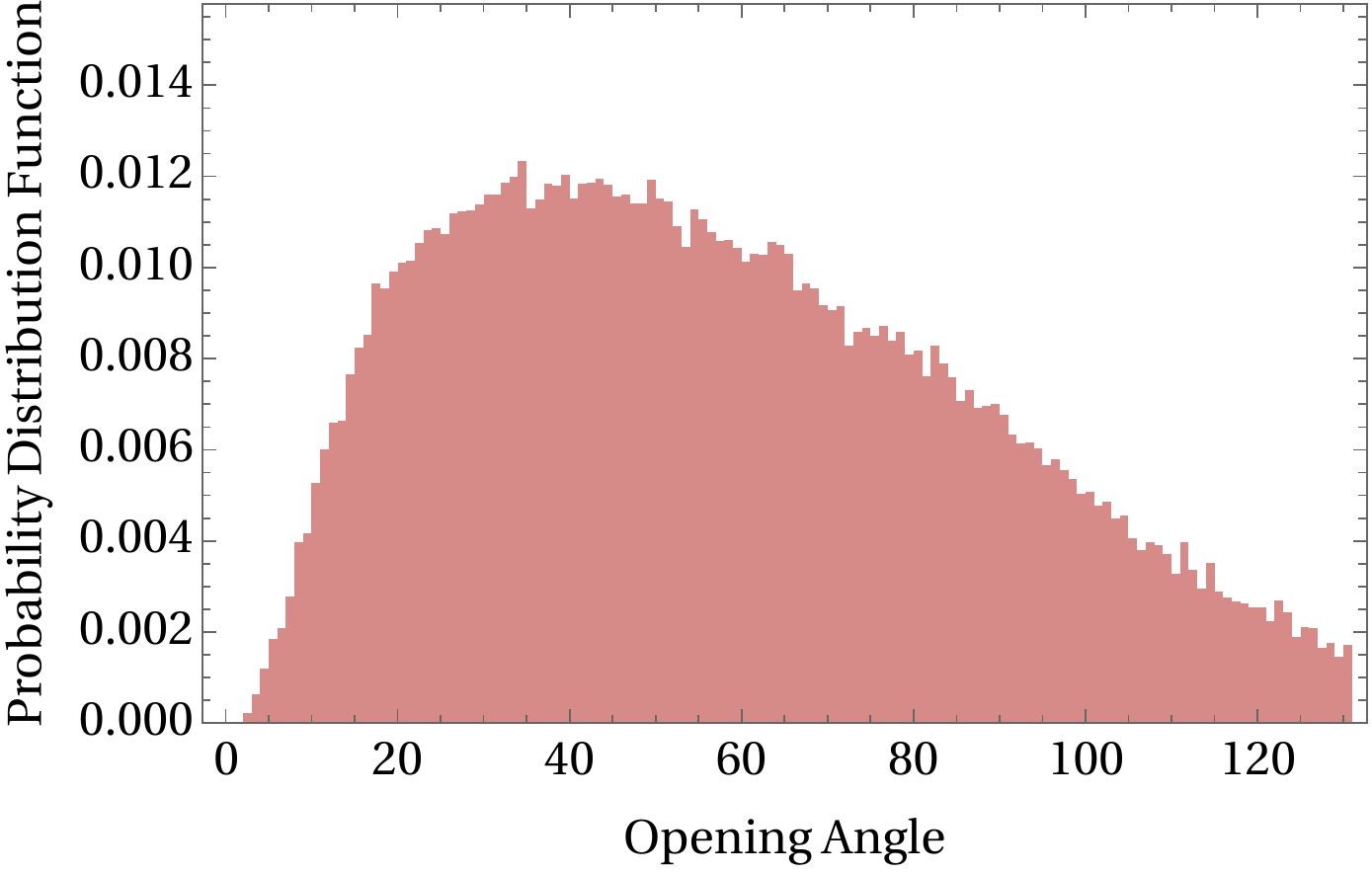}}
\caption{Additional kinematical features of the $\phi$ decay signal. {\bf Left:} Zenith angle $\varphi$ distribution of the incoming into the Hyper-K detector for two sets of parameters, $m_\phi=150\,$MeV, $\theta = 5\times10^{-4}$ (red) and $m_\phi=150\,$MeV, $\theta = 10^{-4}$ (blue). The first point corresponds to the blue star in Fig.~\ref{mainplot} and can leads to hundreds of $\phi$ decay events in Hyper-K. Most of the events are expected to arrive in directions above the detector's horizon.
{\bf Right:} electron-positron opening angle distribution from $\phi$ decay, for $m_\phi=150\,$MeV. The corresponding $\phi$ energy spectrum is shown in Fig.~\ref{fig:fluxes}.
}
\label{fig:kinematics}
\end{figure*}

Moreover, there is important information about the lifetime and mass of $\phi$ in the proposed signal, including the zenith angle and opening angle distributions of the final state $e^+e^-$ pairs. 
In the left panel of Fig.~\ref{fig:kinematics}, we plot the distribution of the zenith angle of $\phi$ particles arriving at the Hyper-K detector, for two sets of parameters.
They exhibit very different behaviors, which can be understood by comparing the $\phi$ decay length, Eq.~(\ref{decaylength}), and the distance it needs to travel before reaching the Hyper-K detector, $L(\varphi)$, given in Eq.~(\ref{Lvarphi}).
The first set of parameters, $m_\phi=150\,$MeV, $\theta = 5\times10^{-4}$, corresponds to the blue fivestar in Fig.~\ref{mainplot}. 
It represents an exciting and uncharted parameter space that will be explored by the upcoming intensity frontier experiments.
In this case, $\gamma \beta \tau_\phi \sim 100\,$km, for a typical boost factor (see Fig.~\ref{fig:fluxes}), whereas $L(\varphi) \sim 10^4,\, 300,\, 10\,$km for $\varphi=0, \pi/2, \pi$, respectively. 
Clearly, if a $\phi$ particle travels to the detector from directions well below the horizon ($0 < \varphi < \pi/2$), the distance $L(\varphi)$ is too long compared to $\gamma \beta \tau_\phi$ for it to survive. As a result, most of the $\phi$ particles are expected to arrive from above the Hyper-K detector's horizon ($\pi/2<\varphi<\pi$).
For comparison, the second set of parameters has a much smaller $\theta$ leading to a much longer lived $\phi$, $\gamma \beta \tau_\phi \sim 10^4\,$km, thus $\phi$ could also arrive from directions below the horizon. 
However, smaller $\theta$ means fewer $\phi$ being produced from the atmosphere and such a point is beyond the reach of Hyper-K. 
Similarly, as $m_\phi$ increases beyond twice of the muon mass, it mainly decays into $\mu^+\mu^-$, via a much larger muon Yukawa coupling. The corresponding decay length is too short for $\phi$ to reach Hyper-K, unless $\theta$ is made much smaller, again resulting in a suppressed atmospheric production rate. In both latter cases, a larger detector would be needed.

In the right panel of Fig.~\ref{fig:kinematics}, we plot the final state electron-positron opening angle distribution from $\phi$ decays,
for $m_\phi=150\,$MeV. The result peaks around $\theta_{e^+e^-} \sim 30^\circ$, which is expected from 
the peak of $\phi$ energy distribution in Fig.~\ref{fig:fluxes}, using
$\theta_{e^+e^-} \sim 2m_\phi/E_\phi$.
A sizable fraction of events have a large ${e^+e^-}$ opening angle. This quantity is relevant for the double ring signature to be resolved
once they occur inside the Hyper-K detector. 
In the main plot Fig.~\ref{mainplot}, we did not implement any cut on $\theta_{e^+e^-}$, which is straightforward to do once the threshold is established.

To summarize, we propose broadening the purpose of the Hyper-Kamiokande experiment though using it to hunt down long-lived Higgs portal scalar particles produced from the atmosphere. 
This proposal is in high complementarity to the intensity frontier experiments exploring rare meson decays.
The target parameter space is for the scalar mass below twice the muon mass that is allowed by existing searches. The corresponding signal is electron-positron pair creations in the Hyper-K detector. 
We make approximations to the atmospheric production picture and derive a semi-analytical expression for the signal rate.
In most events, the electron-positron opening angle is large enough for the double-ring signal to be resolved. If the double-rings are further used to reconstruct the decaying $\phi$ particles, one would find most of $\phi$ are arriving from directions above the detector's horizon. 
In the future, a more inclusive treatment of the $\phi$ production, better understanding of angular distribution measurement by the Hyper-K detector, as well as the background will be useful toward deriving a precise limit.
The Hyper-K reach reported here for Higgs portal scalar similarly applies to light axion-like particles (of the DFSZ type~\cite{Dine:1981rt, Zhitnitsky:1980he}) which couple to Standard Model fermions also through their masses.
The presence of small electron Yukawa coupling in the decay rates naturally makes these particles long lived and suitable to be searched for at earth-sized experiments.

It could be exciting to explore the proposed signal using the existing Super-K data. However, it is worth noting that the Super-K detector volume is about a factor of ten smaller than Hyper-K~\cite{Fukuda:2002uc}. 
One could also consider searching for the signal at the future DUNE far detector which is made of liquid argon and is a few times smaller than Hyper-K in volume.
This said, DUNE could be better at distinguishing $e^\pm$ from $\gamma$, which is useful for background discrimination.
It is beyond the scope of this paper to quantitatively compare the performance between Hyper-K and DUNE.

There have been recent proposals of further searching for long-lived light particles including the Higgs portal scalar at accelerator neutrino facilities using their near detectors~\cite{Batell:2019nwo, Berryman:2019dme, Foroughi-Abari:2020gju, microboone2020}, as well as higher energy collider experiments with displaced detectors~\cite{Beacham:2019nyx} (see the middle panel of Fig.~\ref{mainplot}). In comparison, the atmospheric $\phi$ particles carry relatively lower energies than their beam counterpart,
thus the resulting $e^+e^-$ opening angles are wider and easier for detection. Background is also much lower in the absence of a nearby intense beam. The very large Hyper-K detector volume partially compensates for the relatively lower atmospheric luminosity. All in all, there is excellent complementarity between the searches for long-lived particles of atmospheric and beam origins.

\bigskip

\noindent{\it Acknowledgement.} We thank Razvan Gornea for helpful discussions on the Hyper-K experiment, and Paddy Fox and Roni Harnik for discussions at early stage of this work. Y.Z. is supported by the Arthur B. McDonald Canadian Astroparticle Physics Research Institute.

\bibliography{Atmospheric.bib}
\end{document}